# Power of Mediation Effects Using Bootstrap Resampling

Jason Steffener[*]

Interdisciplinary School of Health Sciences, University of Ottawa, Ottawa, ON Canada

**\* Corresponding author**

Mailing address:

Interdisciplinary School of Health Sciences,

University of Ottawa,

200 Lees, Lees Campus,

Office # E250C,

Ottawa, Ontario, CANADA

K1S 5S9

Phone: 613 562-5800 x4314

# Abstract

Mediation analyses are a statistical tool for testing the hypothesis about how the relationship between two variables may be direct or indirect via a third variable. Assessing statistical significance has been an area of active research; however, assessment of statistical power has been hampered by the lack of closed form calculations and the need for substantial amounts of computational simulations. The current work provides a detailed explanation of implementing large scale simulation procedures within a shared computing cluster environment. In addition, all results and code for implementing these procedures is publicly available. The resulting power analyses compare the effects of sample size and strength and direction of the relationships between the three variables. Comparisons of three confidence interval calculation methods demonstrated that the bias-corrected method is optimal and requires approximately ten less participants than the percentile method to achieve equivalent power. Differing strengths of distal and proximal effects were compared and did not differentially affect the power to detect mediation effects. Suppression effects were explored and demonstrate that in the presence of no observed relationship between two variables, entrance of the mediating variable into the model can reveal a suppressed relationship. The power to detect suppression effects is similar to unsuppressed mediation. These results and their methods provide important information about the power of mediation models for study planning. Of greater importance is that the methods lay the groundwork for assessment of statistical power of more complicated models involving multiple mediators and moderators.

## Introduction

Mediation analysis is a statistical method for investigating how the relationship between two variables is altered once a third variable is entered into the model (Fairchild & MacKinnon, 2009; Hayes & Rockwood, 2017). These models provide the ability to test hypotheses about the explanatory role of how a variable indirectly affects another via a third mediating variable. These types of analyses gained popularity with the framework laid out by Baron and Kenny in their influential paper (Baron & Kenny, 1986). Their method is defined as a qualitative procedure for assessing mediation. Although this provides an excellent framework for interpreting results, it does not provide a formal statistical test analyzing the size of the indirect effect. Furthermore, this approach is considered to have low power and should no longer be followed (Hayes, 2009; David P. MacKinnon, Lockwood, Hoffman, West, & Sheets, 2002).

Limitations of Baron and Kenny's (1986) approach were addressed with the use of the Sobel test (Sobel, 1982) which assessed the statistical significance of the indirect effect. However, this advance for the field was limited by the strong normality assumptions that the Sobel test made (Bollen & Stine, 1990). To improve upon the limitations of the Sobel test, bootstrap resampling approaches have been tested and have been found to provide robust assessments of statistical significance (Efron & Tibshirani, 1994). Specifically, bootstrap resampling procedures provide distributions of values for calculating confidence intervals around indirect effect sizes (Haukoos & Lewis, 2005). Additionally, there are a number of ways to calculate confidence intervals including the percentile method, the bias-corrected (BC) method, and the bias-corrected and accelerated (BCa) method; the use of the BC method is currently recommended in the literature (Hayes, 2018).

Beyond the assessment of the significance of mediating effects, there is a need for evaluation of statistical power (Zhang, 2014) with the use of simulations providing a potential avenue for assessment (Thoemmes, Mackinnon, & Reiser, 2010). When data is simulated using known parameter values, the significance of the mediation effect may be calculated. Repeating this process

...4

many times, and counting the number of significant results, provides a measure of power for chosen parameter values.

Simulation techniques have been used in past research for assessing power. One approach used the Sobel test to assess significance (Zhang, 2014). Therefore, this approach did not account for guidelines that bootstrap approaches are optimal with mediation models. A similar approach has been developed using simulations with bootstrap assessment of confidence intervals; however, this approach used the percentile calculation of confidence intervals (Schoemann, Boulton, & Short, 2017). Unfortunately, this work did not use the recommended BC methods for calculating confidence intervals. Additionally, Schoemann and colleagues (2017) only presented their methods, and did not include their results, which is likely due to the extremely large number of calculations required for assessing power with simulations and resampling.

Work by Fritz and MacKinnon (2007) used simulations to assess statistical power in mediation analyses and compared multiple statistical methods for assessing significance (Fritz & Mackinnon, 2007). The current work has similar foundations and expands on their important work in a number of important manners. First, the methods described here form the basis for assessing power in more complicated path models that include multiple mediators and moderators. Therefore, the use of distributed computing is a key feature of this work and all code is publicly available, see data availability statement below. In addition to providing all code, all results from these simulations are also available. Secondly, the use of bias-corrected and accelerated confidence interval calculations is included which is an expansion on their work. Finally, the current work provides an overview of how proximal and distal effects, suppression effects and collinearity all affect the statistical power of mediating effects.

## Methods

Mediation models use linear regression models to split the total effect that an independent variable (X) has on a dependent variable (Y) into a direct effect and an indirect effect via a third variable (M), see Figure 1. The indirect effect is considered the mediating effect.

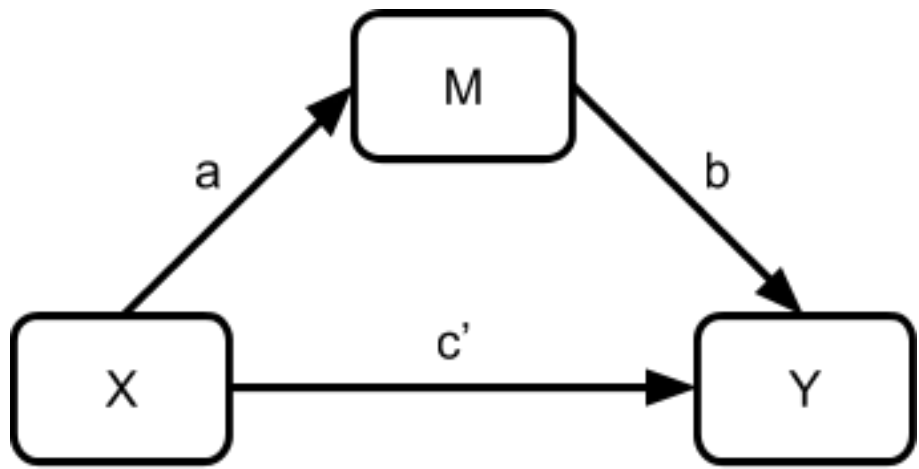

Figure 1. The mediation model used in these simulations. X is the independent variable, Y is the dependent variable, and M is the mediating variable. The effect of X on M is a; the effect of M on Y, independent of X, is b; and the effect of X on Y, independent of M, is c'. Not included in this figure is the total effect of X on Y, which has an effect of c.

The total effect regression model is:

1. $Y_i = cX_i + e_i$

where *i* indicates an individual's value on the dependent or independent variable.

Once the mediating variable is entered into the model, the equation becomes:

2. $Y_i = c'X_i + bM_i + e_i$

The parameter c, in equation 1, is the **total** effect that X has on Y. After including M into the model, c' is the **direct** effect that X has on Y after accounting for M. A third equation is used predicting the mediating variable M with the independent variable X:

3. $M_i = aX_i + e_i$

The indirect effect is then the difference between the total effect and the direct effect, which in turn equals the multiplication of the effects of X on M and of M on Y:

$$c - c' = ab$$

Mediation is assessed for statistical significance through testing of the *ab* effect. Even when these parameters are normally distributed, their product is no longer normal, thereby violating an assumption of the Sobel significance test (Sobel, 1982), making bootstrap significance testing more appropriate.

When the significance of the indirect effect is assessed through bootstrapping, the data is resampled with replacement and the models in equations 2 and 3 are fit to the resampled data and the indirect effect is calculated. From the resultant distribution of values, confidence intervals are determined to identify whether or not they include zero. A result not overlapping with zero is considered significant at the specified alpha value.

In order to determine the power within mediation models, simulated data are generated using predetermined parameter values and sample sizes. The regression models are fit, the indirect effect is calculated and bootstrap resampling is used to estimate significance. This process of simulating data and bootstrap resampling is repeated, and the proportion of times a set of parameter values has a significant result is considered the power. This approach relies on resampling and simulation and not a closed form of calculations. Therefore, this approach allows for power to be assessed in increasingly complicated path models. These would be models that include moderation and moderated-mediation.

To simulate the data, parameter values were predefined for variables a, b and c' through a range of values along with different sample sizes, see Table 1. The total effect, *c*, was then calculated as *c = c' + a\*b*. The simulated data began as three vectors of length N, a mean of one, and a standard deviation of one. The X vector was then weighted by parameter *a* and added to M. This new M vector was weighted by *b* and added to Y along with the X vector weighted by *c.'*

**Confidence Intervals**

Three approaches for calculating confidence intervals were chosen: percentile, bias-corrected, and bias-corrected accelerated. These are listed from least to most complicated to implement. The methods are discussed in the same order since they can be considered nested within each other.

The percentile (PER) method simply defines the confidence intervals as the percentile at each tail. This assumes there is no bias or skew in the resampled distribution with respect to the point estimate. The bias-corrected (BC) confidence intervals correct for any deviations there may be between the center of the bootstrap distribution and the point estimate. For instance, using 1,000 resamples, a bias would be more resamples to the left of the point estimate than the right. The bias-corrected, accelerated (BCa) approach corrects for any bias in the distribution as well as any skew in the distribution with respect to the point estimate. The manner in which these three approaches are implemented is such that in the absence of any bias or skew, the methods produce the same confidence intervals as the more straightforward methods. Therefore, theoretically there is no reason to not use the more complicated BCa approach.

With respect to implementation, the BCa approach involves a jack-knife step, whereas the PER and BC methods simply involve the bootstrap distribution and no significant amount of additional calculations. The BCa however, requires the re-estimation of the regression models an additional N (sample size) times for the jack-knife. The jack-knife performs a leave-one-out approach to the data, thus requiring larger and larger calculations as sample sizes increase (Frangos & Schucany, 1990).

Table 1. Parameter values used for simulations

| Variable | Minimum value | Maximum value | Step size | Total number of values |
|---|---|---|---|---|
| Sample Size (N) | 10 | 200 | 10 | 20 |
| X to M (a) | -0.5 | 0.5 | 0.1 | 11 |
| M to Y (b) | -0.5 | 0.5 | 0.1 | 11 |
| X to Y, in the presence of M (c') | -0.5 | 0.5 | 0.1 | 11 |
| Total scenarios | | | | 26,620 |
| Number of bootstrap resamples | | | | 1000 |
| Number of repeats per simulated scenario | | | | 1000 |
| Regressions per bootstrap resample | 55*2 = 110 (55 is the average sample size across the range used) | | | |
| Total regression model fits | $2.9 \times 10^{12}$ = 2.9 Trillion model fits | | | |

**Implementing simulations**

Performing the simulations required implementation within a high performance computing environment. Resources of the *Cedar* high performance computing facility of ComputeCanada were used. The simulations were organized such that for each scenario, a chosen N, a, b and c', with all of its bootstrap resampling and repeats to estimate power was considered a single job. Each job consisted of approximately 110 million regression model fits. Within the computing environment used for this work, one such job required approximately 36 minutes. This time varied based on the sample size due to the jack-knife procedure. Within the computing environment this would require the submission of 26,620 jobs. To minimize the number of job submissions, job arrays were used.

Due to the nature of the simulations, jobs were created with one parameter being an array of values. Therefore, with arrays, a single submitted job can assess all values for a single parameter. Using one of the parameter weights with 11 different values, reduced the total jobs to be submitted down to 2,420. When using a distributed computing environment, jobs are submitted to a queue to await processing. As the cluster environment is a shared resource, each user is allowed to submit a maximum of 1,000 jobs at one time. Therefore, a single job or 1000 submitted jobs each take approximately 36 minutes to complete. The use of job arrays, therefore allows for 11,000 different simulations to be run at once. In conclusion, with this approach, the nearly three trillion model fits can be completed in two to three hours.

Unfortunately, submitting 2,420 jobs to a computing cluster may not always be successful. Some jobs are not submitted properly due to time-out issues or may encounter random errors. Therefore, this project also implemented housekeeping approaches to ensure all jobs were submitted and successfully completed. The first step created all job submission scripts. These are bash scripts (*.sh files) that call python with the simulation parameters. When these scripts are being written, a table of simulation parameters is created with flags for whether the scenario is complete or not. A second script reads the list of scenarios table and cross checks it against an output folder of results. If results do not exist for an expected scenario, the appropriate bash script is resubmitted to the computing cluster. A counter keeps track of how many jobs have been submitted and stops at the defined user limit, e.g., 1000 jobs per user. Once the jobs have finished in the queue, this step is repeated until all jobs are finished.

# Results

Results demonstrate the power of the indirect effect as a function of parameter and sample sizes. In addition, power is assessed in a number of ways, which include full and partial mediating effects; as a function of the confidence interval calculation method; with respect to differing distal and proximal effect sizes; as a suppression effect and with respect to collinearity. Results are presented here graphically and are also publicly available as a large table. The code used to perform all simulations and to generate all figures is also available.

## Power of the Indirect Effect

The power of the indirect effect was calculated across a wide range of a, b and c' parameter values for sample sizes between 10 and 200. There were no differences in power for positive or negative mediation effects. Positive mediation effects occurred from both a and b parameters being of the same sign. Negative mediation was the result of parameters being of opposite signs. Only the magnitude of the parameter values affected power.

Results show that when one parameter is 0.1, then the indirect effect will not have sufficient power regardless of the size of the other effect, see Figure 2. Similarly, when one effect has a size of 0.2, the sample size needs to be at least 175 participants in order for the indirect effect to have sufficient power. When the effects are both large, sufficient power is achieved with sample sizes below 40.

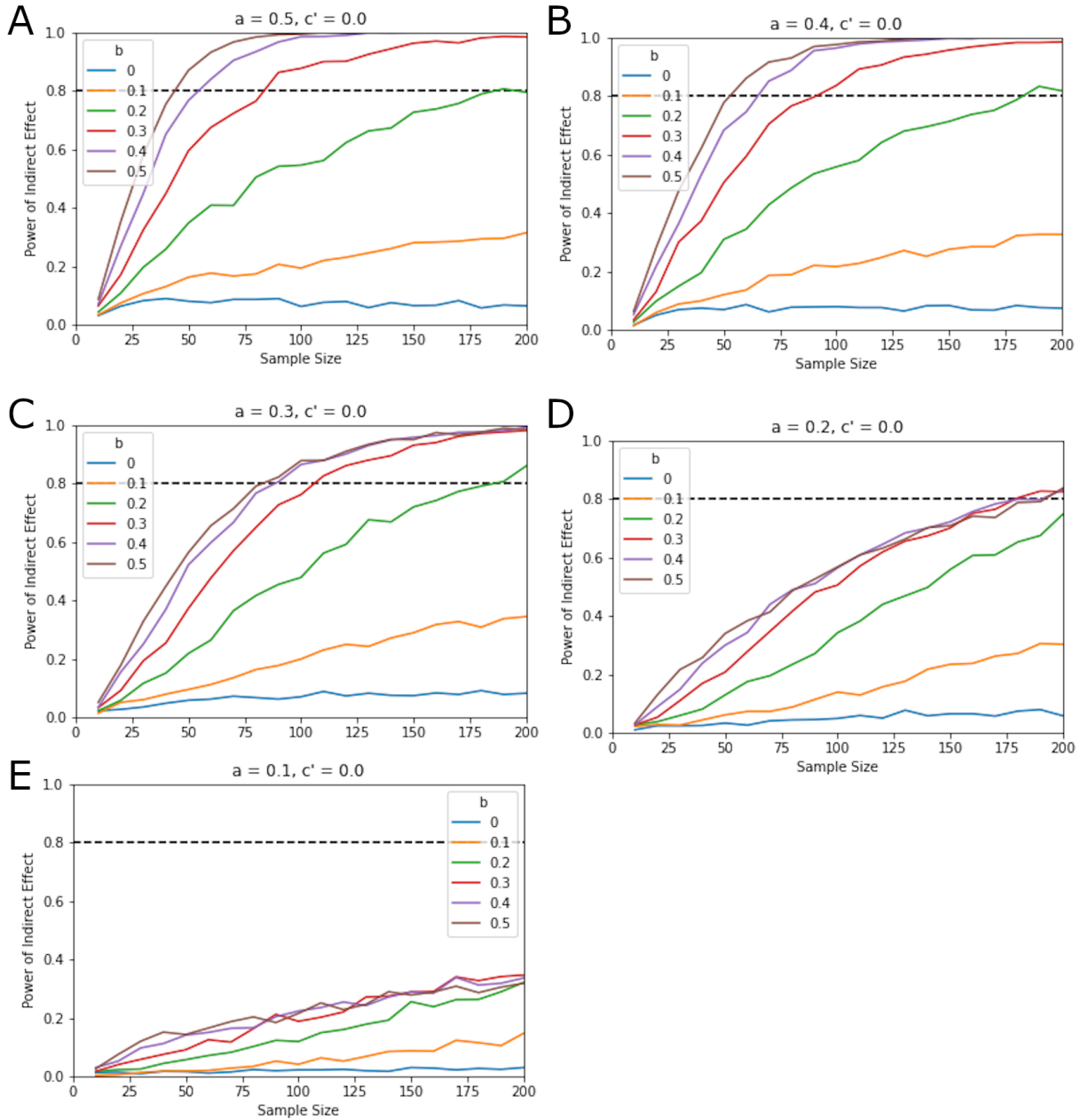

Figure 2. Power of the indirect effect. Each plot shows the power of the indirect effect on the vertical axis versus sample size along the x-axis. All plots have the same direct effects (c') of zero and show all simulated values b (0, 0.1, 0.2, 0.3, 0.4, 0.5). The horizontal dashed line represents power of 80%. A) the a effect is 0.5, B) the a effect is 0.4, C) the a effect is 0.3, D) the a effect is 0.2, E) the a effect is 0.1.

**Power of the Indirect Effect in the Presence of Partial to Full Mediation**

Full mediation describes the situation where there is a significant total effect and a non-significant direct effect. Therefore, the mediator is considered to fully explain the relationship between the dependent and independent variables (Baron & Kenny, 1986). Partial mediation describes the situation when the direct effect is significant. Therefore, the mediator is considered to only partly explain the relationship between the dependent and independent variables.

Results were explored to determine whether there were differences in power depending on whether the mediator fully or partially accounted for the relationship between the dependent and independent variables. Choosing one value for both *a* and *b* parameters, plots of power for various values of *c'* were generated. Power of the mediating effect was not affected by the size of the unmediated effect of A on C. These results, Figure 3, also demonstrate that the power of the mediation effect is not affected by the size of the total effect.

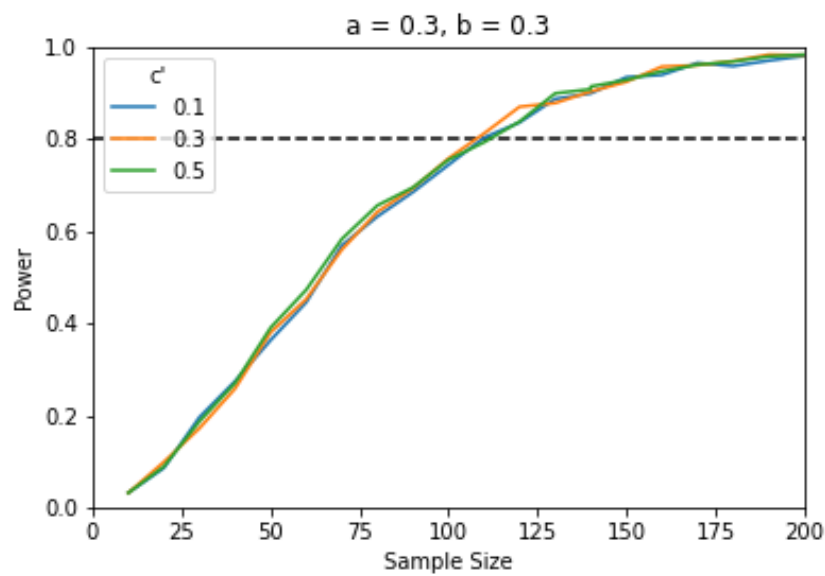

Figure 3. Power of indirect effect as a function of partial and full mediation. The power of the indirect effect is shown using fixed values of parameters a and b, each set to 0.3, with varying sizes of the direct effect, c'.

**Difference in Power Across Different Confidence Interval Calculation Methods**

Choosing one value for *a, b* and *c'* parameters, plots of power for the three confidence interval methods were generated. These plots demonstrate similar results for both methods that account for bias and lower power for the percentile method that does not account for bias or skew in the bootstrap resample results. Figure 4, plots the power versus sample size for the situation where *a* and *b* are both set to have values of 0.3. For power of 0.8 an additional 9 participants are required for the percentile method to equal the power of the bias corrected methods..

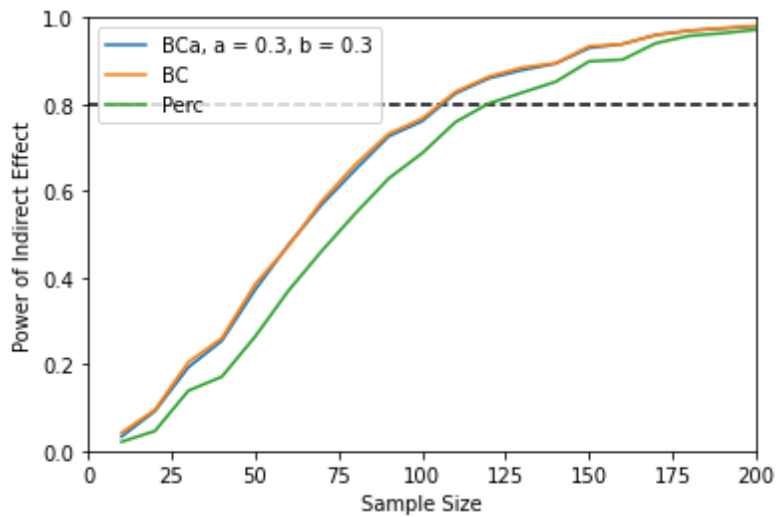

Figure 4. Power of the indirect effect as a function of the confidence interval calculation method. The power of the indirect effect is shown using fixed values of parameters a and b, set to values 0.3, with three methods of calculating confidence intervals. BCa: bias-corrected, accelerated confidence intervals which adjust for both bias and skew in the bootstrap resamples. BC: bias-corrected confidence intervals which adjust only for bias. Perc: percentile method which does not make any corrections on the bootstrap resample distribution.

**Proximal and Distal Effects**

Proximal and distal effects refer to the size of the effects in the B to C arm or the A to B arm, respectively. Within an experimental framework where the temporal order of effects was A to B and B to C, with C being the most recent, it is expected there to be greater effect sizes in the proximal relationship as compared to the distal one. The current approach cannot assess the actual temporal order with simulated data; however, it can assess whether there are statistical findings that also support this idea. There were no observed differences in power for mediation effect sizes when the distal and proximal effects varied. Figure 5 demonstrates matched estimated power across samples sizes for greater proximal or greater distal effect sizes.

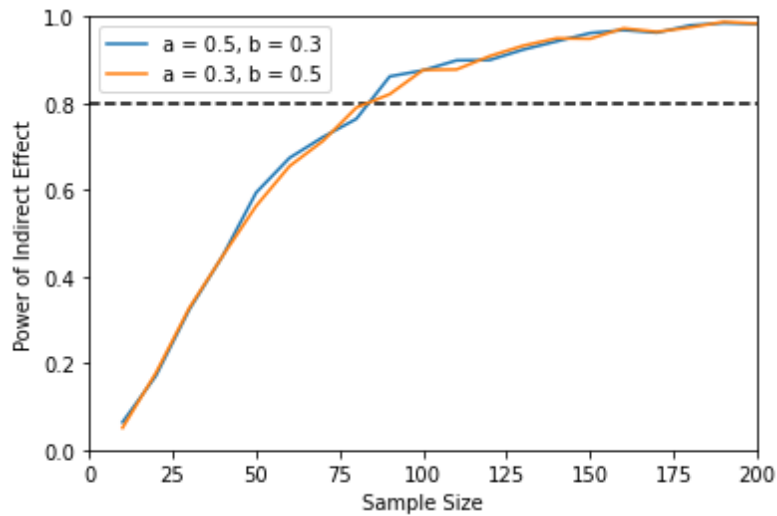

Figure 5. Proximal and distal effects. The power for the indirect effect in the presence of differing sizes of the proximal and distal arms of the model.

**Suppression**

One major limitation to the Baron and Kenny approach is that it requires there to be a significantly large total effect, A has to significantly predict C. It is argued that with a test of the size of the mediating effect a*b, this restriction may be dropped. When this restriction is dropped, it also allows for exploration of suppression effects. A suppression effect is when there is no observed total effect of A on C; however, there is a significant indirect effect (D. P. MacKinnon, Krull, & Lockwood, 2000).

Suppression effects occur when the relationships between A to B and B to C are in opposite directions resulting in a negative indirect effect. The remaining direct effect is positive and together they cancel each other out producing a null total effect. This is referred to as suppression since once the mediator is entered into the model, the relationship between A and C is revealed, as a direct effect, where it was suppressed in the absence of the mediator in the model. Suppression effects can also occur when the mediating effect is positive; however, the direct effect is negative. Figure 6 shows these three scenarios where all demonstrate sufficient power of the indirect effect for sample sizes above 100 even when there is no total effect in the data.

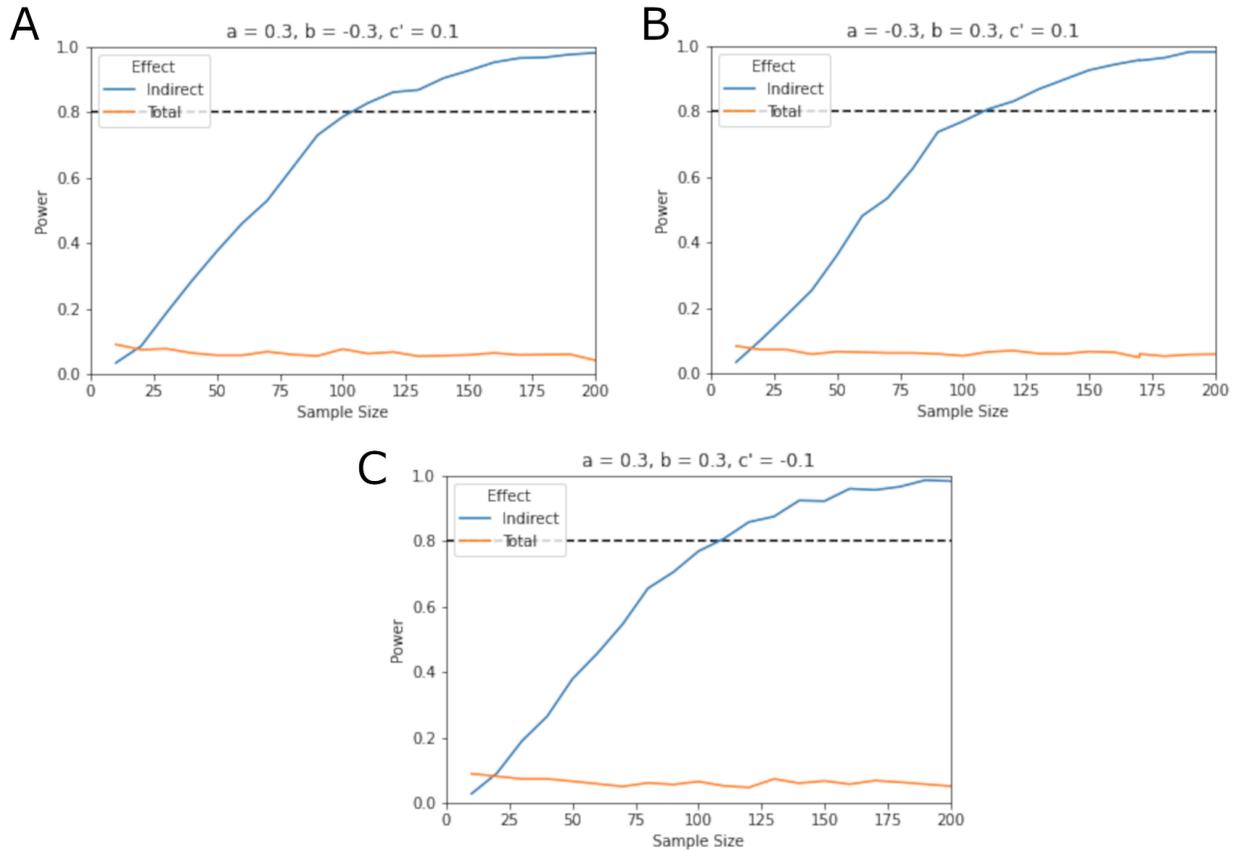

Figure 6. Suppression Effects. Three scenarios demonstrating suppression effects where the total effect is near zero; however, there exists an indirect effect. This occurs when the direction of the indirect and direct effects are in opposite directions and similar magnitude. In all three scenarios there is increasing large power for the indirect effect as the sample size increases with no increase in power for the total effect. A) the A to B pathway has an effect of 0.3, the B to C pathway has an effect of -0.3 and A to C, after accounting for B is +0.1. B) the A to B pathway has an effect of -0.3, the B to C pathway has an effect of 0.3 and A to C, after accounting for B is 0.1. C) the A to B pathway has an effect of 0.3, the B to C pathway has an effect of 0.3 and A to C, after accounting for B is -0.1.

**Collinearity**

As the strength of the relationship between X and M increases, there is an inherent increase in the collinearity between predictors when modelling Y. Collinearity between predictors in a model impacts the ability to detect significant results. To explore the impact of collinearity, the power of the direct effect of X on Y was explored as the size of the effect of X on M was varied. Simulation results demonstrated that the power of the direct effect did in fact differ as a function of the size of a, the effect of X on M. The effect however is relatively small within the range of effect sizes used , see Figure 7.

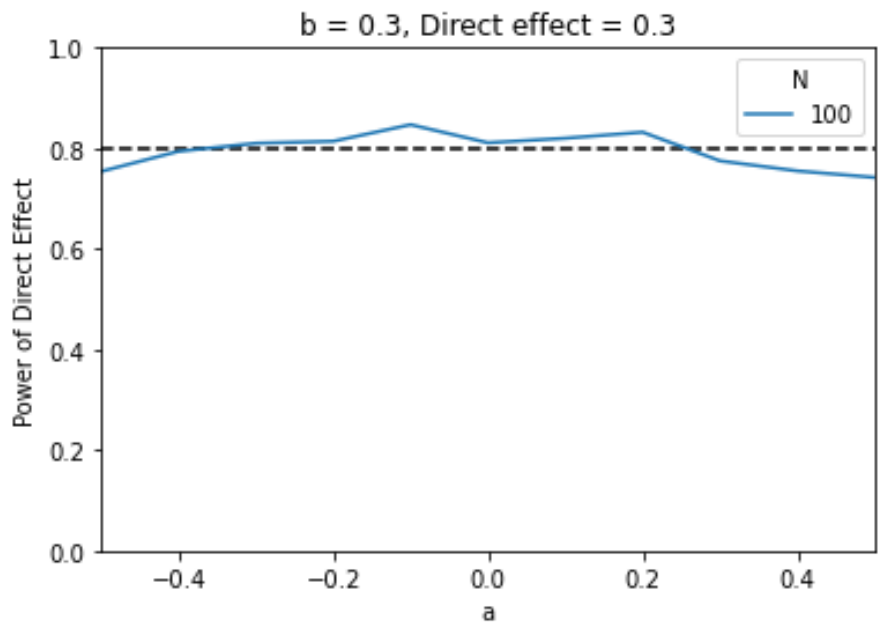

Figure 7. Collinearity within a mediation model. In this scenario the B to C pathway, accounting for A has an effect of 0.3 and the direct effect (the A to C pathway, accounting for B) is 0.3. With a sample size of 100, as the A to B effect varies from large negative to positive values, the power of the direct effect decreases relatively small amounts at the extreme values when collinearity is the greatest.

## Discussion

This work provides estimates of the achieved power for varying sample sizes and parameter values for mediating effects. Cohen's guidelines for effect sizes as small (0.14), medium (0.39) and large (0.59) (Cohen, 1988) are used to interpret these results. Due to the nature of the simulated data with standard deviations of one, the chosen parameters are standardized effect sizes. When the effect sizes for either branch of the mediating model are small, regardless of the size of the other branch, the mediating effect does not have sufficient power up to a sample size of 200. When the effect size of either branch is medium, the other branch needs to be at least 0.2 (between small and medium) to detect a significant mediating effect with a sample size of approximately 180. Medium effect sizes for both branches require a sample size of approximately 100. A large effect size for both branches requires a sample size less than 50 for sufficient power. These interpretations are based on using bias-corrected confidence intervals.

The choice of confidence interval does impact the power of mediating analyses. The percentile method requires a sample size that is approximately 10 participants greater than the two methods that include bias corrections. In these simulations, no improvement was found when including bias as well as skew adjustments. This is an important finding due to the additional complexities and computational requirements when correcting for skew in the bootstrap distributions. Estimating skew in the bootstrap resamples requires jack-knife resampling which estimates the model as many times as there are samples in the data as well as bootstrap resampling. The similarity in confidence intervals between both approaches means the computationally simpler approach provides equivalent power estimates as the more statistically and computationally sophisticated bias-corrected, accelerated method. This is an finding which will be incorporated into the application of mediation analyses with brain imaging data which apply mediation analyses across hundreds of thousands of data points throughout the brain (Steffener, Barulli, Habeck, & Stern, 2014; Steffener, Gazes, Habeck, & Stern, 2016)

      The Baron and Kenny (1986) method for mediation requires the presence of a significant total effect before testing for mediating effects. This makes logical sense by implying that there needs to be a relationship present between two variables before one can explore what explains the relationship. However, the bootstrap methods for assessing statistical significance in mediation do not require this assumption. Therefore, it is possible to identify the presence of significant indirect effects between variables via a mediator when there is no total effect between variables in the absence of a mediator. The simulations demonstrate such results and highlight the fact that the power of an indirect mediating effect is independent of the size of the total effect.

      Collinearity is an inherent part of mediation models. The size of the mediation effect is dependent on there being a relationship between variables X and M; however, as this relationship strengthens the collinearity increases when predicting Y. Within the range of effects sizes used in this study, the results demonstrated minimal impact on power in the presence of increasing collinearity. These results differ from the work by Beasley (2014) where large collinearity effects were shown (Beasley, 2014). This difference is likely due to differing ranges of parameter values used. The current work explored power within small to large effects sizes. The most dramatic negative effects of collinearity were demonstrated in Beasley's work at much larger effect sizes. Therefore, we are demonstrating that collinearity is indeed a statistical concern and that it did not raise any practical concern with the range of effect sizes used.

      A significant benefit of the approach taken in this work is its scalability, since the methods used rely on simulations and resampling. Thus, they are applicable to any mediation, moderation or moderated-mediation model, regardless of the complexity of the model. Additionally, use of this approach may finally begin to fill the gap in our knowledge about the statistical power requirements needed to detect significant effects in complicated path models; these include the presence of

multiple mediators in serial or parallel. In addition, these methods allow exploration of the power of moderating effects on indirect effects. Such models include assessment of mediated-moderation to identify the mechanisms underlying interaction effects as well as moderated-mediation.

The methods presented combine both simulations and resampling. The result is an extremely large number of computations required for assessing power. This computational requirement is a limiting factor for others to utilize the proposed methods. Therefore, the methods used in the current work and all calculated results are publicly available providing insight for future study planning.

The current work is not without limitations. It is possible to simulate the data using different approaches. This could be selection of data values from multi-normal distributions. This approach would be applicable with simple mediation models; however, it would not be applicable to models including moderation terms. The current approach of adding weighted variables together (Fritz & Mackinnon, 2007) could be extended to multiplication of variables to produce moderating effects. In addition, the current work used normally distributed variables; oftentimes, variables may have different distributions. This is especially true in aging and lifespan studies where age is uniformly distributed and age-group is dichotomous. Future work will address these investigative avenues.

Overall, the current work is of importance for two main reasons. Firstly, it provides valuable information about the statistical power of mediating effects across a range of sample sizes and effect sizes. Secondly, the current work provides open-access to the results as well as the code. The code demonstrates a method for distributing the simulation processes across large-scale computational networks. Coupled with the provided scheduling and process checking tools are a host of future directions with path models.

## Acknowledgements

This research was enabled in part by support provided by Westgrid (https://docs.computecanada.ca/wiki/Cedar) and Compute Canada (www.computecanada.ca).

## Data Availability

The datasets generated during the current study are available in the Center for Open Science repository, https://osf.io/5tnv4/. The software used for these analyses are available on GitHib, https://github.com/NCMlab/MediationPowerPaper.

## Financial Support

The authors did not receive support from any organization for the submitted work.

## Conflicts of Interest

The authors have no relevant financial or non-financial interests to disclose.